\documentclass[a4paper, amsfonts, amssymb, amsmath, preprint, showkeys, nofootinbib, twoside]{revtex4-1}
\usepackage[english]{babel}
\usepackage[utf8]{inputenc}
\usepackage[colorinlistoftodos, color=green!40, prependcaption]{todonotes}
\usepackage{soul}
\usepackage{amsthm}
\usepackage{mathtools}
\usepackage{physics}
\usepackage{xcolor}
\usepackage{graphicx}
\usepackage[left=23mm,right=13mm,top=35mm,columnsep=15pt]{geometry} 
\usepackage{adjustbox}
\usepackage{placeins}
\usepackage[T1]{fontenc}
\usepackage{lipsum}
\usepackage{csquotes}
\usepackage[pdftex, pdftitle={Article}, pdfauthor={Author}]{hyperref} 
\usepackage{algorithm}
\usepackage{algpseudocode}
\begin{document}
\title{Soft aperture spatial filtering: 1.5W in a single spatial mode from a highly multi-mode laser diode in an external cavity}
\author{Mallachi-Elia Meller, Idan Parshani, Leon Bello, David Goldovsky, Amir Kahana, Avi Pe'er}
    \email[Correspondence email address: ]{avi.peer@biu.ac.il}
     \affiliation{Department of Physics and BINA Institute of Nanotechnology, Bar-Ilan University, Ramat-Gan 52900, Israel}
    
\date{\today} 

\begin{abstract}
Broad area laser diodes are attractive for the high optical power they can produce. Unfortunately, this high power normally comes at the cost of severely reduced spatial coherence since the wide area of the semiconductor wave-guide is inherently spatially multi-mode (in the slow axis). We demonstrate a method to majorly improve the spatial coherence of a high-power broad-area diode laser by placing it in an external cavity that is mode selective. We design the cavity, such that the diode aperture acts as its own spatial filter, obviating the need for an intra-cavity slit-filter, and optimally utilizing the entire gain medium. We demonstrate this soft filtering method using wide diodes of $200 \rm{\mu m}$ and $300 \rm{\mu m}$ widths and compare its power-efficiency to the standard approach of hard-filtering with a slit. We obtain high-gain operation in a pure single-mode, demonstrating up to $1.5\rm{W}$ CW power at $1064 \rm{nm}$ with high beam quality.  
\end{abstract}


\maketitle

\section{Introduction}
Semiconductor lasers are advantageous over other types of gain media: in their high electrical-to-optical efficiency, simple and robust construction, high optical power, tunability, wide range of available wavelengths that cover the entire VIS-NIR range, and low cost. Consequently, high-power semiconductor lasers with a good beam quality (brightness) are desirable for many applications, such as pump sources for solid-state lasers and fiber lasers \cite{applications1,applications2}, frequency conversion \cite{applications3,applications4}, direct material processing \cite{applications5,applications6} and medical applications \cite{applications7}. However, these two requirements, brightness and power, do not normally go hand-in-hand \cite{bpm1}. In order to increase the output power, while avoiding nonlinearities and saturation effects, it is necessary to increase the gain-medium area. This requires wide wave-guides, which are inherently multi-mode. Consequently, the number of spatial modes increases, which severely degrades the brightness of the diode and its spatial coherence.

The standard method for overcoming this problem in broad aread amplifier (BAA) diodes is to try to enforce a single spatial-mode on the diode laser by placing it in an external cavity \cite{external1,external2,external3} that stimulates the BAA with spatially selective feedback. The spatial selectivity is normally achieved by introducing a slit at the Fourier-plane of the diode facet \cite{slit1,slit2}, which acts as a spatial filter to determine the number of allowed spatial modes and their width. Unfortunately, this hard-aperture method often endures unnecessary losses that hamper its power-efficiency due to the inherent mismatch between the lowest spatial mode of the diode wave-guide and that of the slit. 

We suggest to mitigate this limitation using a ``soft filtering'' design of the cavity, where the diode itself acts as the spatial filter, which eliminates any possible mismatch from the source. While some "soft filtering" cavity designs were demonstrated in the past in an anamorphic external cavity \cite{anamorphic1} , the obtained power was relatively low ($120\rm{mW}$ in single mode). We demonstrate soft filtering at much higher power levels with much wider diodes, obtaining up to $1500\rm{mW}$ in a single mode beam.

\section{Mode-selective cavity design - Soft filter vs Hard Slit}
\label{sec:methods}
\begin{figure}[h!]
    \centering
    \includegraphics[width=0.8\textwidth]{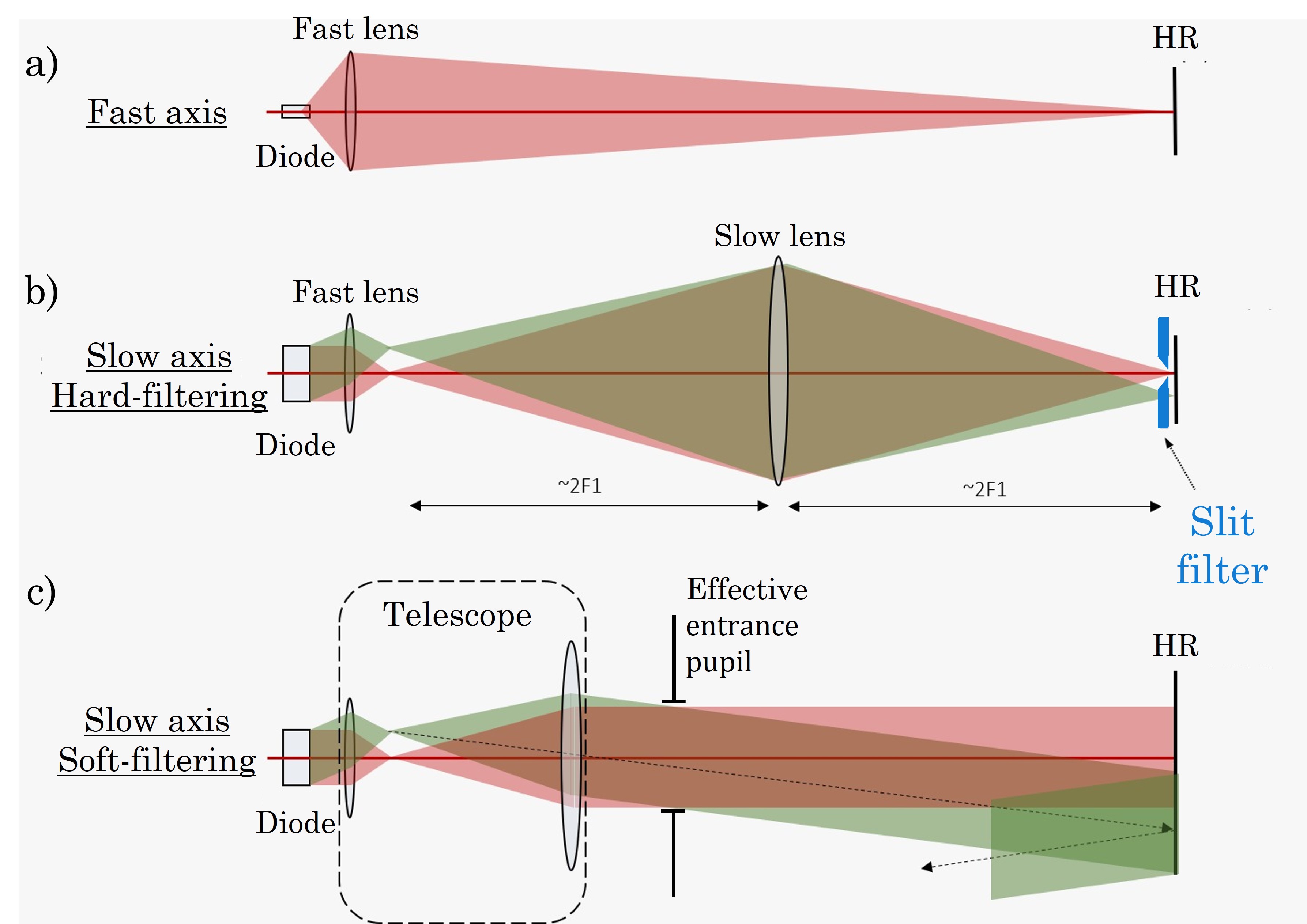}
    \caption{Cavity beam shaping mechanism \textbf{(a)} \textbf{Fast-axis:} The fast axis is inherently single-mode due to the diode structure, in order stabilize the fast axis inside the cavity, a spherical fast lens is located near the diode at a distance slightly larger than its focus. This forms a magnified, low divergence image of the diode's facet in the fast axis on the end mirror. \textbf{(b)} \textbf{Slit-filtering:} The fast spherical lens forms a Fourier transform of the diode's facet plane at its back focal plane. This Fourier plane is then imaged by the cylindrical slow lens onto the end mirror, where the slit is placed. The two colors visualize two different modes of the cavity, showing the slit's operation to filter only the desired mode.
    \textbf{(c)} \textbf{Soft-filtering:} In the soft-filtering approach, we operate in a plane configuration, employing the diode's aperture as its own spatial filter. We shift the position of the slow lens near the telescope position (together with the fast lens), where the diode's magnified image acts as an entrance-pupil for the returning beam from the high-reflector. Single mode operation is enforced when the two-way distance from the effective pupil to itself exceeds the Rayleigh range of a single mode beam with the pupil's width.}
    \label{fig:filtering}
\end{figure}

To introduce mode-selectivity that prefers single-mode operation in the laser cavity over multi-mode, while preserving most of the multi-mode power, we implemented both the standard hard slit-filtering and our modified soft-filtering, as schematically shown in figure \ref{fig:filtering}. To compensate for the highly astigmatic beam at the diode output facet, we employ a common anamorphic design of the cavity, based on two lenses - a spherical lens with a short focus (fast lens) that operates on both the fast and the slow axes of the diode, followed by a cylindrical lens of a longer focus (slow lens) that operates only on the slow axis. The fast axis, which is single-mode by definition, is simply imaged by the fast lens onto the end mirror (figure \ref{fig:filtering}\textbf{a}). The configuration of the slow axis for standard hard-filtering with a slit is shown in figure \ref{fig:filtering}\textbf{b}, and our soft filtering approach is shown in figure \ref{fig:filtering}\textbf{c}. 

To implement a hard slit filter, the fast lens generates a Fourier-transform of the diode aperture at its back focal plane. This Fourier plane is then imaged by the slow lens to the end mirror, where a slit is positioned to filter a single spatial frequency (corresponding to a single plane wave at the diode) to provide feedback to the cavity. The multi-mode operation can be described as an incoherent sum of plane waves that propagate at different directions out of the diode, and are focused onto different points on the end mirror. The slit then selects a range of plane waves at the end mirror, where the width of the slit selects the number of allowed modes. This enforces a wide single-mode at the diode facet that covers most of the diode's aperture. The magnification ratio of the two images on the end mirror (fast and slow) allows to compensate for the spatial astigmatism of the beam. As mentioned, the main problem with slit-filtering is the mismatch between the mode determined by the slit, and the mode determined by the diode, as will be demonstrated in the results section.

In contrast, the soft filtering approach moves the slow lens away from the imaging condition, towards the fast lens, which gradually reduces the accepted beam divergence at the diode plane, until finally only the fundamental mode with the lowest divergence can survive (a single plane-wave). This condition is reached when the slow-lens and the fast-lens form a telescope on the slow-axis, where the image of the diode acts as an effective entrance pupil for the beam in the return path (see figure \ref{fig:filtering}\textbf{c}), which indicates that only a beam with sufficiently low divergence can safely pass the aperture in the backwards direction. When the separation of those two "aperture locations" approaches, or exceeds the Rayleigh range of a single-mode beam with a waist that matches the aperture width, only this lowest spatial mode can survive. Our filtering mechanism operates therefore between two spatial stability limits - the hard slit-filtering in figure \ref{fig:filtering}\textbf{b} is equivalent to a concentric cavity and the soft filtering configuration of figure \ref{fig:filtering}\textbf{c} is equivalent to a planar cavity.

\section{Results}
\begin{figure}[h!]
    \centering
    \includegraphics[width=0.8\textwidth]{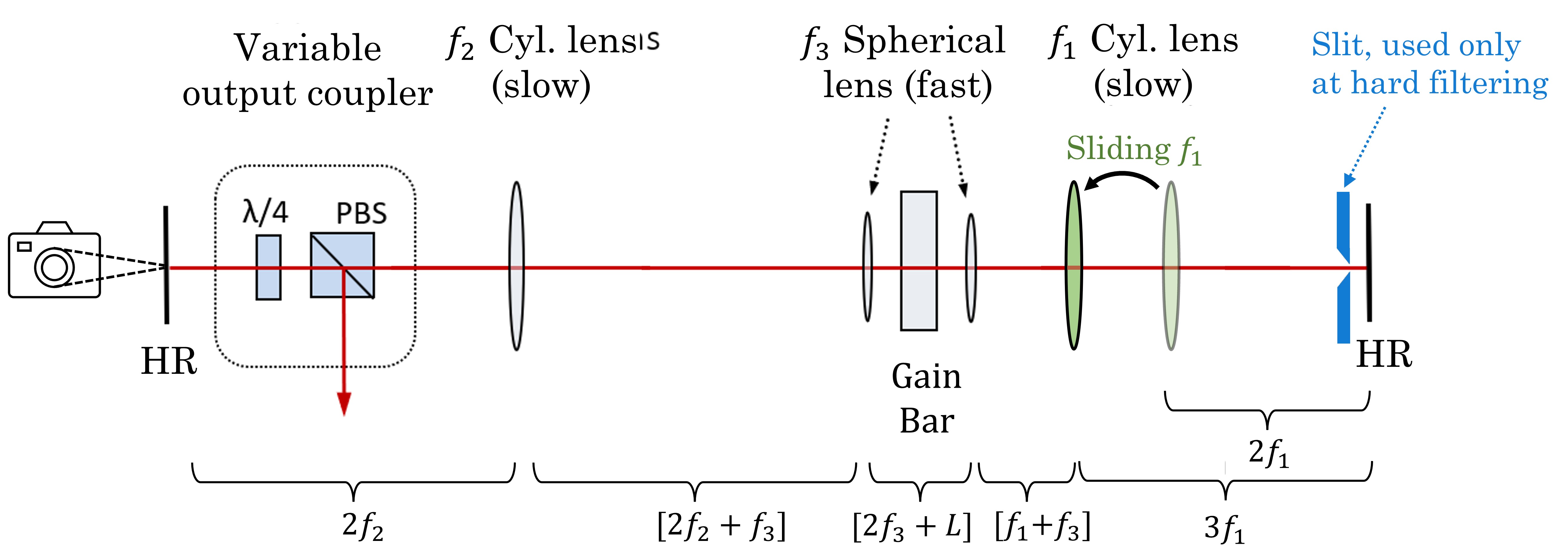}
    \caption{\textbf{Experimental setup:} The construction of the cavity is based on a standard linear cavity where the diode (gain bar) is located in the center. In order to stabilize the fast and slow axis, the diode facet is located roughly at the focal plane of the two spherical fast lenses (f3). The right cavity arm contains the spatial shaping mechanism - for hard-filtering a slow cylindrical lens (f1) is placed at a one-to-one imaging distance of $2f_1$ from the end mirror and from the Fourier-plane of the diode (at the back focus of the fast lens). The slit is applied near the right end mirror. For the soft-filtering the slit is removed, and the slow lens is shifted towards the fast lens. The left cavity arm is multi-mode by design and contains a variable output coupler in the form of a polarized beam splitter and a $ \lambda/4$ wave plate. A CCD camera behind the left HR ($R\approx99.5\%$) is used to analyse the laser beam}
    \label{fig:System-diagram}
\end{figure}

To demonstrate the advantage of soft-filtering over hard filtering we used the experimental cavity configuration of figure \ref{fig:System-diagram}, which is comprised of two arms around the diode - the right arm, which implements the spatial filtering (on both approaches) and a left arm, which supports all modes, and includes a variable output coupler, implemented by a polarizing beam-splitter and a rotating $ \lambda/4$ wave-plate that optimizes the output power (see caption of figure \ref{fig:System-diagram} for details). The left side is operated in the concentric configuration where the diode plane (near field) is Fourier-transformed onto the end mirror (far field). We therefore expect that at the transition from multi-mode to single-mode the beam profile at the left end mirror will converge into a narrow profile of minimum width accompanied by a maximal brightness, as illustrated in Fig. \ref{fig:output_vs_slit}\textbf{(a)}. 

We employed in our experiments two broad-area diodes, to study low-power and high-power regimes: One diode of $200 \rm{\mu m}$ width and $4 \rm{mm}$ length, of relatively low power (up to $220\text{mW}$ multi-mode). Both diodes were AR coated to suppress self-lasing. The low power cavity was comprised of slow lenses with focal length: $\rm{f1} = 50 \rm{mm}$ (right arm, filtering) and $\rm{f2} = 80 \rm{mm}$ (left arm, output coupling) and the fast lenses focal length was $\rm{f3}=7.5 \rm{mm}$. The high-power configuration (up to 2.6W multi-mode) used a larger diode with dimensions $300 \rm{\mu m} \times 5 \rm{mm}$. In addition to the AR-coating, the $300\mu m$ diode was also placed at a small angle to the optical axis to suppress even further self-lasing from reflections of the end-facets. The focal lengths of the lenses in the high-power cavity were $\rm{f1} = 50 \rm{mm}$, $\rm{f2} = 100\rm{mm}$, and $\rm{f3} = 9\rm{mm}$.

\subsection{Hard (slit) filtering}

To explore the performance of hard-slit filtering in detail we first employed the lower power, $200 \rm{\mu m}$ diode in the configuration of figure \ref{fig:System-diagram} and measured the output power and the beam properties for varying slit widths, aiming to quantify the beam purity and the filtering power-efficiency. The diode was operated at a working current of $2.2\rm{A}$, which is sufficiently above lasing threshold of the external cavity ($\approx1.7$A) to generate a highly multi-mode beam, but safely below the threshold for self-lasing of the gain chip alone ($\approx2.6\rm{A}$) due to residual reflections from the diode facets.

\begin{figure}[h!]
    \centering
    \includegraphics[width = 0.6\textwidth]{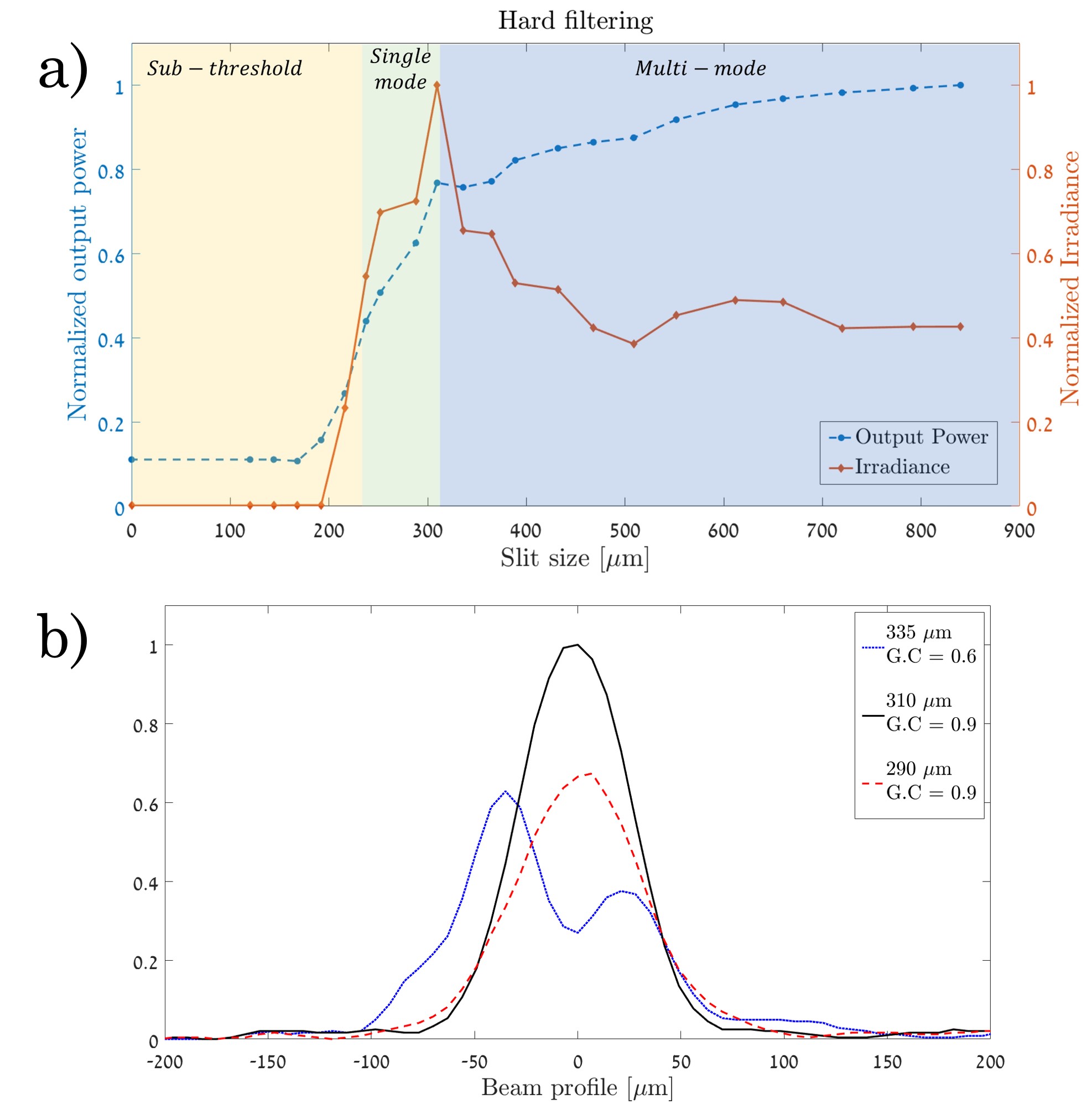}
    \caption{\textbf{(a)} Normalized output power and the normalized irradiance as a function of the slit size, the dashed line is the output power and the solid line is the irradiance measured on the cavity end mirror. For a fully open slit ($850 \rm{\mu m}$) the output power is $160 \rm{mW}$ and the peak irradiance is $40 \frac{\rm{W}}{\rm{mm}^2}$ at $310 \rm{\mu m}$. \textbf{(b)} Beam profiles for three different slit sizes around the transition to single-mode operation.}
    \label{fig:output_vs_slit}
\end{figure}

In the experiment we measured the output-power and the beam profile at the left end-mirror (using a CCD camera) for a varying slit width, as shown in figure \ref{fig:output_vs_slit}. For each slit width we optimized the output coupler, to assure the maximal achievable power-efficiency. Single mode and multi-mode can be distinguished by examining the dependence of the beam irradiance (peak intensity) and the total power on the slit width. The goal of the method is to reduce the number of lasing modes. In spite of the slight reduction of the output power as the slit is closed, the irradiance rises considerably towards the single-mode condition, since the power is spread over less modes, as shown in Fig. \ref{fig:output_vs_slit}(a). 

Looking more closely at the beam properties, we measured the beam profile along the slow axis near the single-mode transition, as shown in figure \ref{fig:output_vs_slit}(b). We observe the crossover point where the slit was sufficient to enforce single mode at $310 \rm{\mu m}$ (solid black). This is the optimal point, where the slit eliminates the additional modes (Gaussian content $=0.9$ \cite{GC1}), while maintaining reasonable mode matching with the diode, such that the output power at single-mode is $>160$mW, which is $\approx75\%$ of the multi-mode power at 2.2A. Closing the slit further (dashed red) remains single mode (Gaussian content $=0.9$), but reduces the power due to increased losses from the slit also for the single-mode. With a wider slit, an additional mode arises, forming a dual-peaked beam profile (solid blue), reducing the Gaussian content to $<0.6$. 

\subsection{Soft-filtering}

\begin{figure}
    \centering
    \includegraphics[width = 0.6\textwidth]{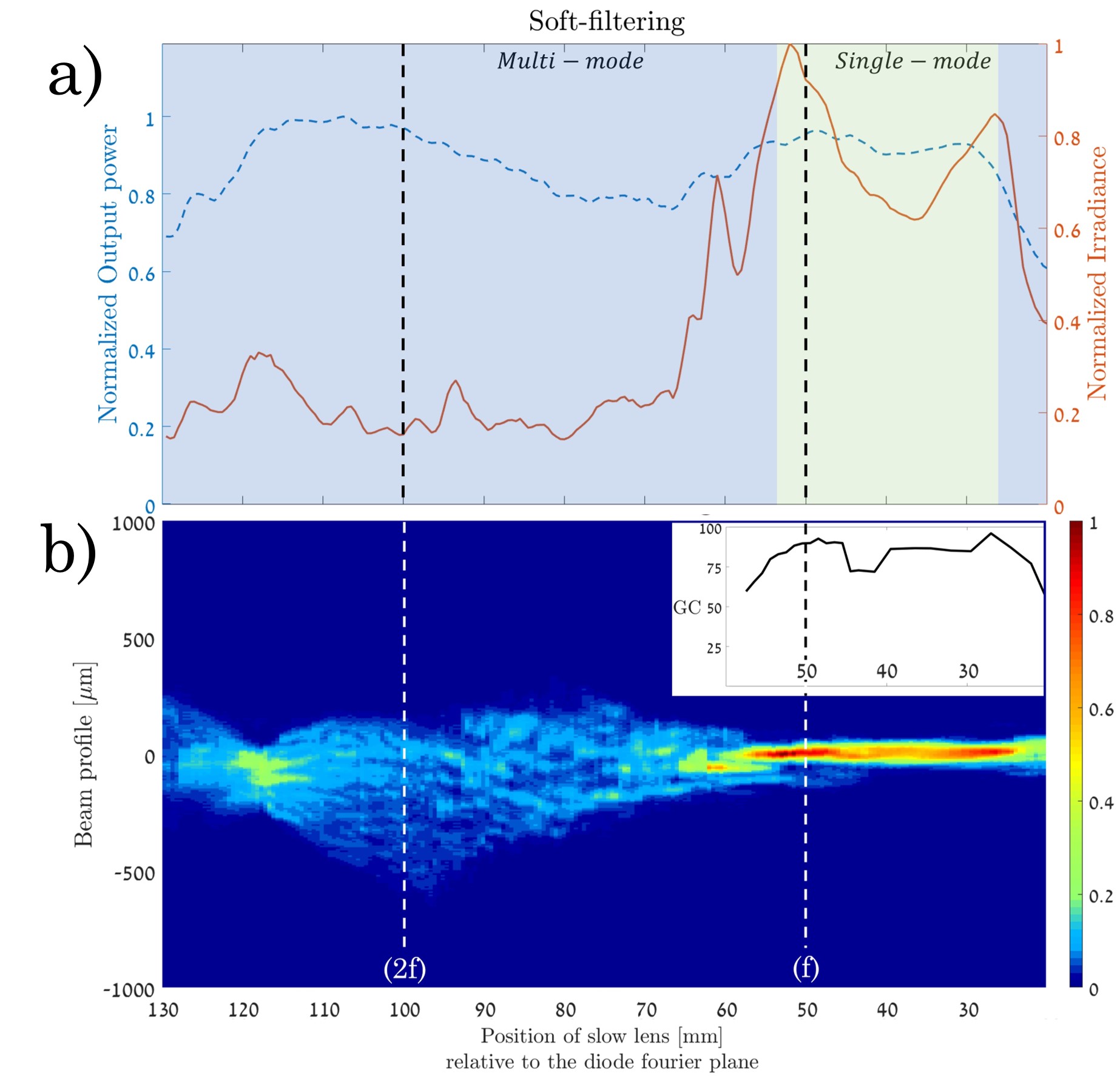}
    \caption{Laser preferments as a function of the cylindrical lenses' (F1) position ($x_{f}$). \textbf{(a)} The left axis (blue dashed curve) is the normalized power, the value at the peak ($x_{f}=92.5 \rm{mm}$) is $166 \rm{mW}$. The right axis (solid orange line) is the normalized irradiance, with peak value ($\rm{x_{f}} = 54 \rm{mm}$) of $68 \frac{\rm{W}}{\rm{mm}^2}$. \textbf{(b)} Line section of the spatial beam profile as a function of the cylindrical lenses' position (f1) position, the $2\rm{f}$ marker is the concentric configuration -- this supports multi-mode operation as can be seen from the spread of the beam at the point. The $\rm{f}$ marker is the plane configuration and as can be seen from this point the beam profile collapses to a single high-brightness point, in the top right corner a measurement of the Gaussian content across the single mode zone.}
    \label{fig:line_profile}
\end{figure}

To demonstrate the soft-filtering method we shift the position of the slow (cylindrical) lens away from the imaging condition. Figure \ref{fig:line_profile}a shows the output power and output irradiance and figure \ref{fig:line_profile}b shows a line scan of the beam profile along the slow axis. both as a function of the position of the slow lens (relative to the back focal plane of the fast lens). The vertical dashed black lines mark the $2f$ imaging location, where the beam is highly multi-mode and the telescope location $f$, where the beam forms a nice single-mode with high Gaussian content ($\approx0.9$), with power that is nearly the same as the maximum multi-mode power ($96\%$). This is a clear advantage over the hard filtering that showed only $75\%$ power efficiency at maximum.

\begin{figure}[h!]
    \centering
    \includegraphics[width = 0.6\textwidth]{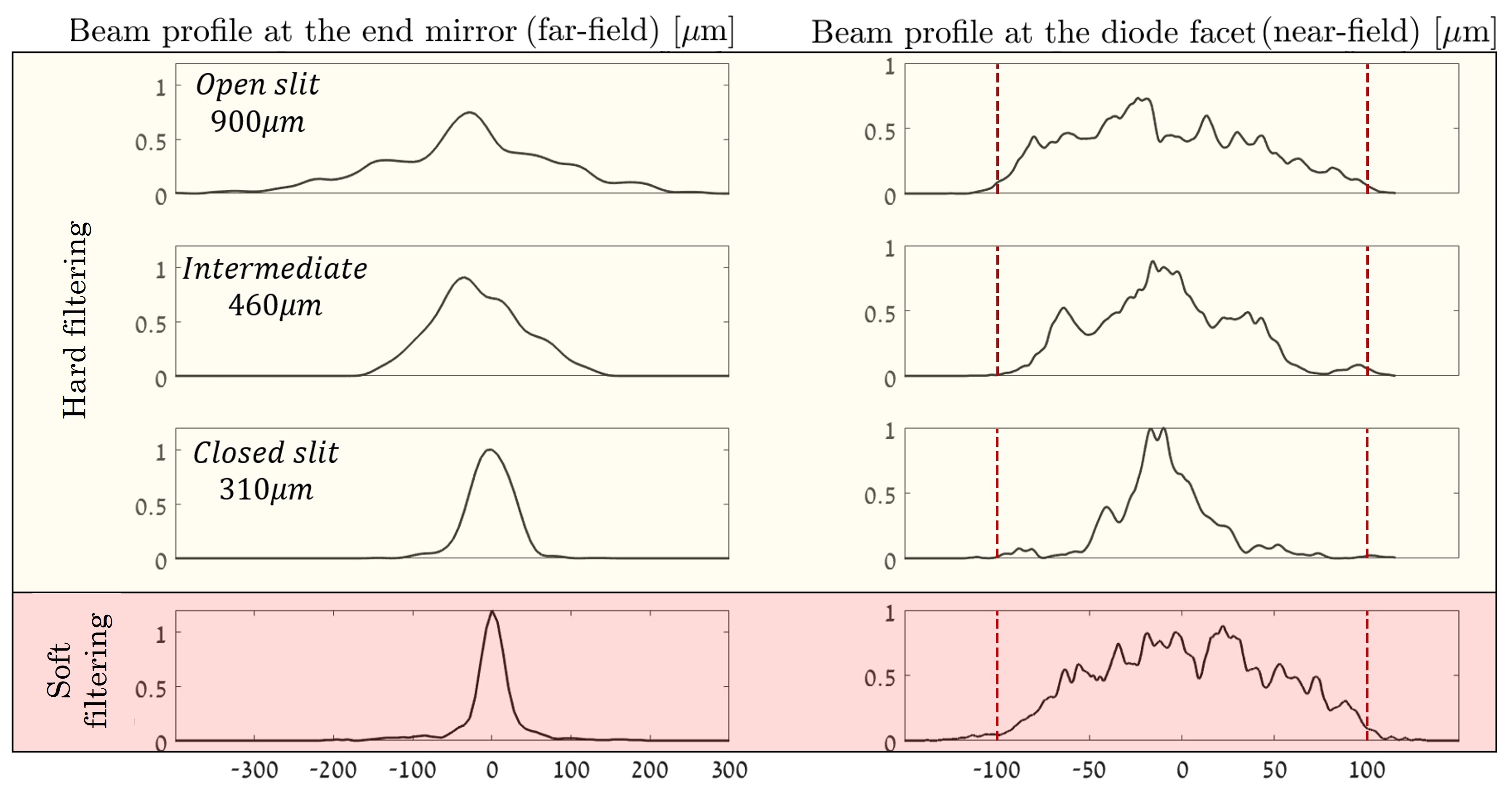}
    \caption{Line section of the spatial beam profile across the slow axis as measured at the far field on the end mirror (\textbf{left column}) and at the near field on the diode facet (\textbf{right column}). The first three rows represent hard-filtering of different slit widths -- \textbf{(i)} completely open, \textbf{(ii)} partially open and \textbf{(iii)} optimum for single mode operation. Row \textbf{(iv)} shows single-mode operation using the soft-filtering configuration.}
    \label{fig:beam_profile}
\end{figure}

The superiority of the soft-filtering is also reflected when examining the beam fill-factor inside the diode, as shown in Fig. \ref{fig:beam_profile} that presents the beam profile on the diode facet (near field) and at the Fourier plane (far field) on the left end-mirror for both hard filtering (top three lines for different slit widths) and soft filtering (bottom fourth line). The physical borders of the diode wave-guide are marked on the near-field profiles by vertical red lines. The best possible fill-factor is represented by the top row, which shows multi-mode operation with no filtering restrictions on the laser operation. Ideally, the single-mode operation should reach a similar fill-factor. Clearly, the hard-filtering is not optimal in this regard, with a fill factor of roughly $50\%$, whereas for the soft-filtering, the fill factor is nearly ideal. This difference is also reflected in the far field beam, where the soft-filtering profile is narrower than that of the hard-filtering, which highlights the fact that the slit mode for hard filtering does not perfectly match the fundamental mode of the diode wave-guide.

\begin{figure}[h!]
    \centering
    \includegraphics[width = 0.6\textwidth]{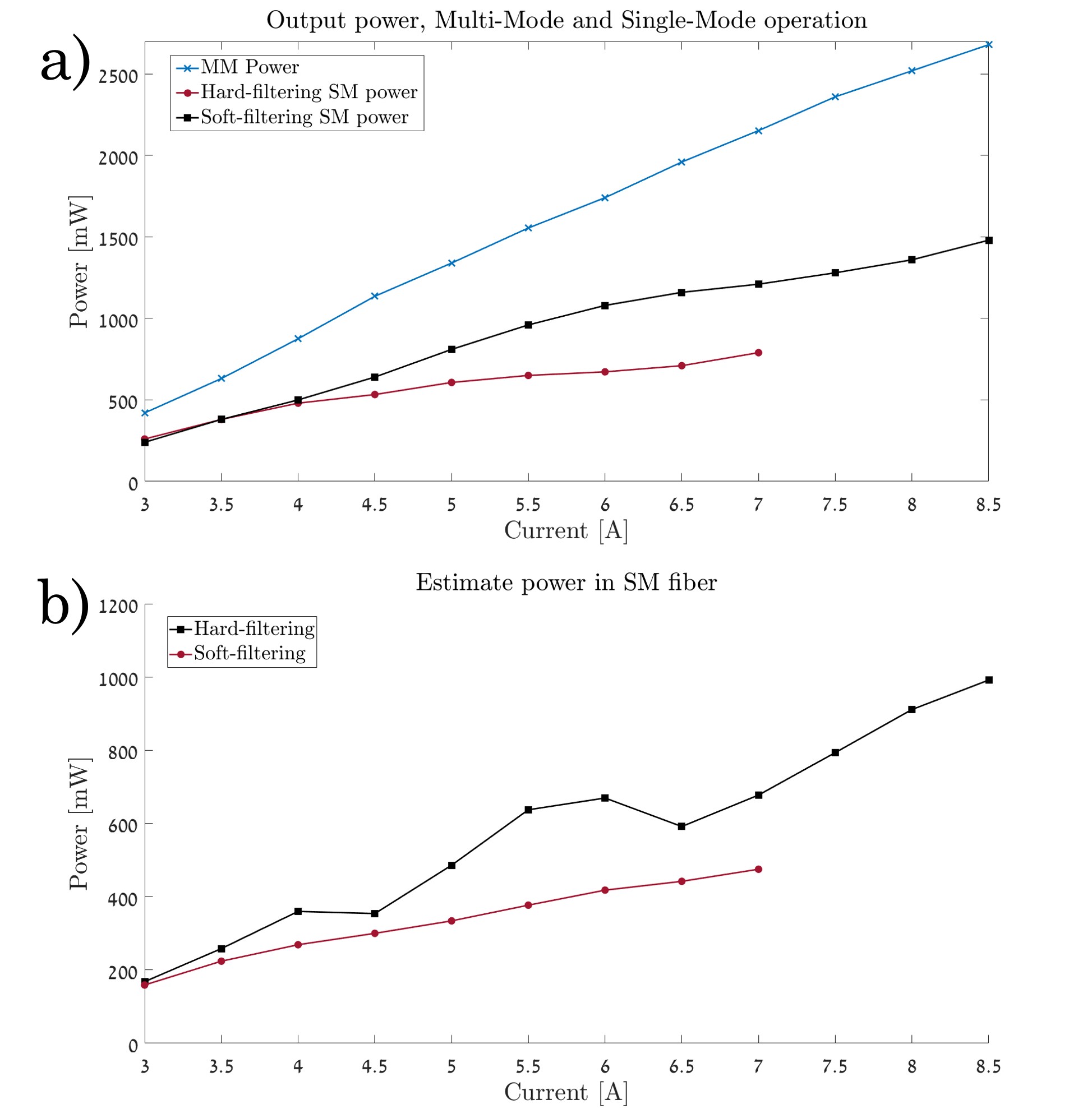}
    \caption{Laser performance in soft and hard filtering. \textbf{(a)} Power-Current curves at single mode operation, the black line is the power obtained using soft-filtering and the red is the power obtained using hard-filtering, the blue line is the output power in multi-mode operation at the confocal configuration. \textbf{(b)} Estimated coupling power to single mode fiber, black line the power received by soft-filtering and the red is the power received by hard-filtering.}
    \label{fig:laser_preferments}
\end{figure}

\subsection{High-power results}
All the results presented thus far were obtained with the narrower diode of $200\mu m \times 4mm$ dimensions, which could be driven only up to $\approx2.6A$ before self-lasing occurred, generating $220\rm{mW}$ of output power at $2.2\rm{A}$. To explore spatial filtering at higher power levels we introduced a wider and longer diode ($300\mu m \times 5mm$ at $1064\rm{nm}$) into the experimental configuration. This wide diode can be driven by much higher currents ($>10\rm{A}$), however, to operate at such high currents, annihilation of self-lasing is absolutely necessary, and only the AR coating is not sufficient for this. To further suppress self-lasing, the wide diode was placed at an angle to the optical axis, which directs the residual reflection out of the diode's wave-guide and prevents it from amplifying. We then applied both the hard and soft-filtering approaches to the high-power wide diodes and examined their performance and limitations. 

Figure \ref{fig:laser_preferments}a shows power-current (PI) curves of the wide diode in multi-mode (blue), as well as the optimal single mode configurations for both hard (black) and soft (red) filtering. Figure \ref{fig:laser_preferments}b presents the estimated power into a single-mode fiber based on the calculated Gaussian content of the output beam. Clearly, the soft filtering performed better than hard filtering in two aspects: First, the overall power efficiency was near $60\%$ for soft filtering, noticeably improved compared to hard-filtering. Second, with soft filtering the diode could be driven to higher currents, while maintaining single-mode operation. Specifically, the maximum current with soft filtering was $8.5\rm{A}$, which generated $1.5\rm{W}$ of output power at single-mode, compared to $<1\rm{W}$ at $7\rm{A}$ for hard filtering. 

\begin{figure}[h!]
    \centering
    \includegraphics[width = 1\textwidth]{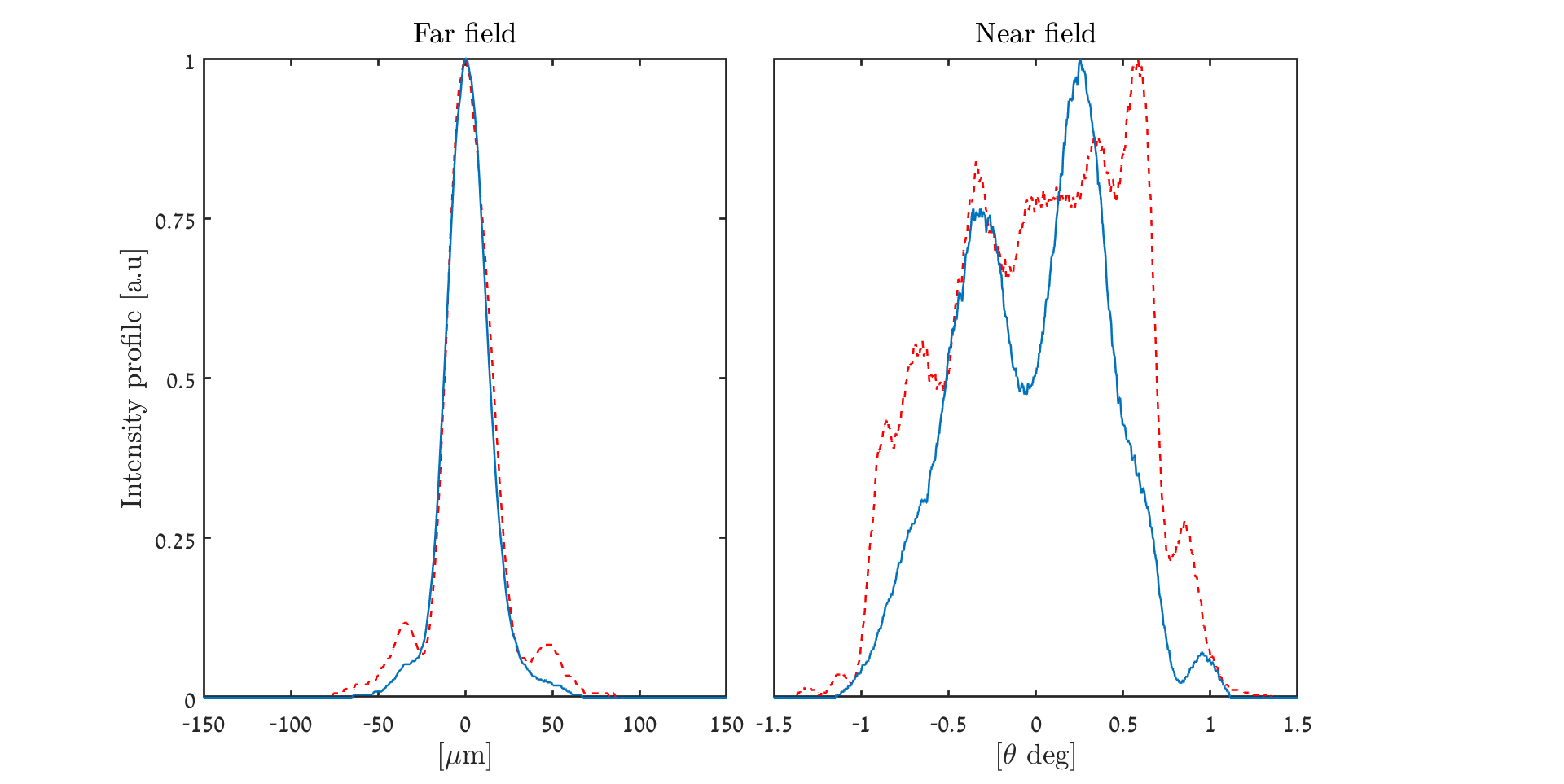}
    \caption{Beam intensity profile normalized by the peak intensity. \textbf{Near field - (left):} the beam intensity profile as seen on the end-mirror. \textbf{(Far field - (right):} beam intensity profile, as imaged outside the cavity. \textbf{Red dashed lines} indicate the raw data, \textbf{solid blue lines} indicate the data with the side-lobes filtered out in the near field using a slit. $M^2$ value for the unfiltered beam was $1.9$, and $1.3$ for the filtered beam.}
    \label{fig:M2_preferments}
\end{figure}

Finally we measured the beam quality of the single-mode beam for both soft and hard filtering using two parameters - $M^2$ and Gaussian content. The Gaussian content for all drive currents was between $0.7-0.8$, and $M^2<1.5$ was obtained when low intensity side lobes and image noise (contain less than 10$\%$ percent of the total power) were omitted ($M^2<2$ for the raw beam image) as demonstrated in figure \ref{fig:M2_preferments} for $8.5\rm{A}$.  

\section{Conclusions}
\label{sec:conclusions}
We demonstrated soft spatial filtering in broad-area, high-power semiconductor lasers, where the diode aperture itself acts as the spatial filter in an external cavity configuration. We extracted up to $1.5\rm{W}$ of high-brightness power, with $1\rm{W}$ power that can be efficiently coupled to a single mode fiber. We demonstrate an improvement in the single mode power and irradiance compared to state-of-the-art methods of hard slit-filtering by roughly $50\%$. Another advantage of our soft-filtering method is the ability to operate at higher working currents, without being negatively affected by self-lasing. Shaping the beam inside the cavity efficiently enforced single-mode operation in a very high-gain regime, which is highly desirable for many applications in the field of laser engineering. 


\bibliography{main.bib}
\end{document}